\documentclass[3p]{elsarticle}
\usepackage{amsmath,amssymb,epsfig,color,mathrsfs,bbm,subcaption}
\usepackage{graphicx}

\begin{document}

\title{How (non-) linear is the hydrodynamics of heavy ion collisions?}


\author{Stefan Floerchinger$^1$, Urs Achim Wiedemann$^1$\\
Andrea Beraudo$^{1,2}$, Luca Del Zanna$^{3,4,5}$, Gabriele Inghirami$^{3,4}$, Valentina Rolando$^{6,7}$
}

\address{$^1$Physics Department, Theory Unit, CERN, CH-1211 Gen\`eve 23, Switzerland\\
$^2$ Dep. de Fisica de Particulas, U. de Santiago de Compostela, E-15782 Santiago de Compostela, Galicia-Spain\\
$^3$ Dipartimento di Fisica e Astronomia, Universit\`a di Firenze, Via G. Sansone 1, I-50019 Sesto F.no (Firenze), Italy\\
$^4$ INFN - Sezione di Firenze, Via G. Sansone 1, I-50019 Sesto F.no (Firenze), Italy\\
$^5$ INAF - Osservatorio Astrofisico di Arcetri, L.go E. Fermi 5, I-50125 Firenze, Italy\\
$^6$ INFN - Sezione di Ferrara, Via Saragat 1, I-44100 Ferrara, Italy\\
$^7$ Dipartimento di Fisica e Scienze della Terra, Universit\`a di Ferrara, Via Saragat 1, I-44100 Ferrara, Italy
}

\begin{abstract}
We provide evidence from full numerical solutions that the hydrodynamical evolution of initial density fluctuations in heavy ion collisions can be understood order-by-order in a perturbative series in deviations from a smooth and azimuthally symmetric background solution. To leading linear order, modes with different azimuthal wave numbers do not mix. Quadratic and higher order corrections are small and can be understood as overtones with corresponding wave numbers.
\end{abstract}


\maketitle

In recent years, fluid dynamic simulations of relativistic heavy ion 
collisions have provided strong evidence for a picture according to which the momentum distributions of soft hadrons result from a fluid dynamic evolution of initial density 
fluctuations, see Refs.~\cite{Heinz:2013th,Gale:2013da,Hippolyte:2012yu,Teaney:2009qa} for recent reviews. The research focusses now on understanding in detail the mapping from fluctuations in the initial state to experimentally accessible observables in the final state~\cite{Qiu:2011iv,Schenke:2011bn,Bhalerao:2011yg,Schenke:2012wb,Holopainen:2010gz,Teaney:2010vd,Teaney:2012ke,Gardim:2011xv,Petersen:2012qc,Qian:2013nba,Niemi:2012aj,Deng:2011at}.
The present letter aims at quantifying to what degree this hydrodynamic mapping is linear in the strength of initial fluctuations around some suitably chosen background, and on what scale non-linearities arise. This is of interest since an {\it approximately} linear relation (by which we mean a mapping in which non-linearities can be understood as small corrections of a predominantly linear mapping) 
would provide a particularly simple and thus particular powerful 
tool for relating experimental observables to the initial conditions of
heavy ion collisions and to those properties of matter that govern their fluid dynamic 
evolution~\cite{Floerchinger:2013rya}. 

We consider initial conditions of heavy ion collisions, specified
in terms of fluctuating fluid dynamic fields $h_i$ on a hyper surface 
at fixed initial time $\tau_0$. Here, the index $i$ runs over all independent fields,
\begin{equation}
	h_i(\tau,r,\varphi, \eta) =
 \left(w, u^r, u^\phi, u^\eta, \pi_{\text{bulk}},  \pi^{\eta\eta}, \dots \right)\, , 
\label{eq1a}
\end{equation}
including e.g. the enthalpy density $h_1=w$, three independent fluid velocity 
components, the bulk viscous tensor, the independent components of the shear 
viscous tensor, etc. In the following we assume Bjorken boost invariance and drop the rapidity-argument $\eta$ in the hydrodynamical fields. Following Refs.~\cite{Floerchinger:2013rya,Floerchinger:2013vua},
we express $h_i$ in terms of a background component $h^{BG}_i$ and an appropriately normalized perturbation $\tilde h_i$.
The background is taken to be a solution of the non-linear hydrodynamic
equations initialized at $\tau_0$ with an azimuthally symmetric
average over many events. It is evolved with the fluid dynamic solver 
ECHO-QGP\cite{DelZanna:2013eua}.
For any sample of events, this background needs to
be determined only once. The time evolution of the 
$\tilde h_i$ is viewed as a perturbative series on top of the background fields,
\begin{eqnarray}
&& \tilde h_i(\tau, r, \varphi) =  \int_{r^\prime, \varphi^\prime} \mathscr{G}_{ij}(\tau, \tau_0, r, r^\prime, \varphi-\varphi^\prime) \;
\tilde h_j (\tau_0, r^\prime, \varphi^\prime)  \nonumber \\
&& \qquad + \frac{1}{2} \int_{r^\prime, r^{\prime\prime}, \varphi^\prime, \varphi^{\prime\prime}} \mathscr{H}_{ijk} (\tau,\tau_0,r,r^\prime, r^{\prime\prime}, \varphi-\varphi^\prime, \varphi-\varphi^{\prime\prime}) \;
\tilde h_j(\tau_0, r^\prime, \varphi^\prime)\;
\tilde h_k(\tau_0, r^{\prime\prime}, \varphi^{\prime\prime})
 + \mathcal{O}(\tilde h^3)\, ,
\label{eq1}
\end{eqnarray}
where $\int_r = \int_0^\infty dr\, r$, $\int_\varphi = \int_0^{2\pi} d\varphi$ etc.
The kernels $\mathscr{G}_{ij}$, $\mathscr{H}_{ijk}$ (and 
corresponding terms for higher orders in $\tilde h_i$) depend
on the time-evolved background $h^{BG}_i$ only. 
Due to the azimuthal rotation symmetry of the background, 
$\mathscr{G}_{ij}$ depends on the angles $\varphi$ and $\varphi^\prime$ 
only via the difference $\varphi-\varphi^\prime$ and similarly for $\mathscr{H}_{ijk}$. 
The question we raise in
the title can now be made more precise: We ask whether the 
expansion (\ref{eq1}) is possible for a suitably chosen background~\footnote{Hydrodynamic evolution is governed by non-linear partial differential equations and it may be chaotic or it may contain terms that are non-analytic in the initial fluid fields $\tilde h_j$. Hence, the validity of the expansion (\ref{eq1}) is not guaranteed. Also, it will
depend on the choice of the background $h^{BG}$ and on the strength of the
perturbations $\tilde h$.} and whether it is dominated by the first linear term. 
To address this question, we compare in the following numerical results from a
full causal dissipative hydrodynamic evolution to expectations based on 
the structure and on the symmetries of the perturbative series (\ref{eq1}).

\begin{figure}[t]
\begin{center}
\includegraphics[width=7.5cm]{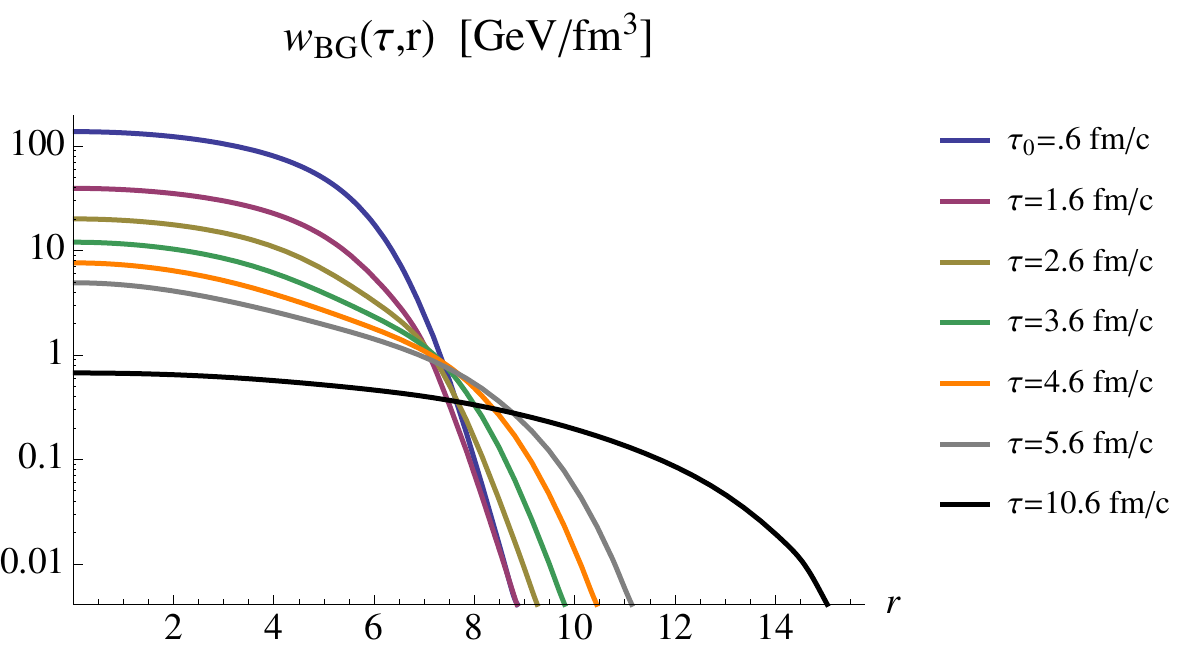}
\includegraphics[width=7.5cm]{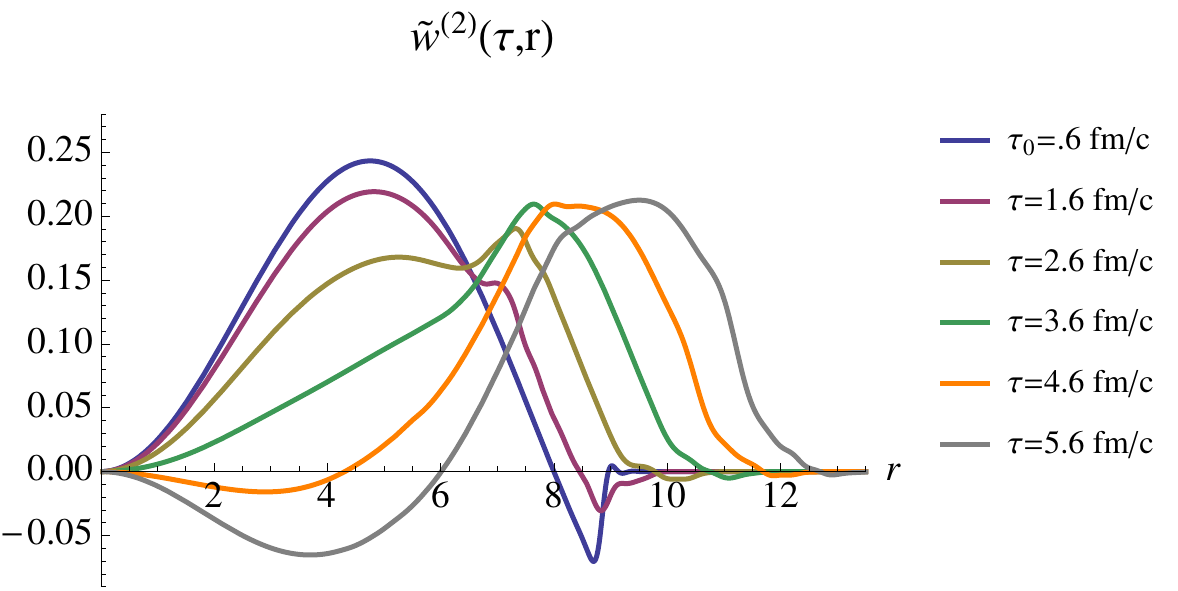}
\includegraphics[width=7.5cm]{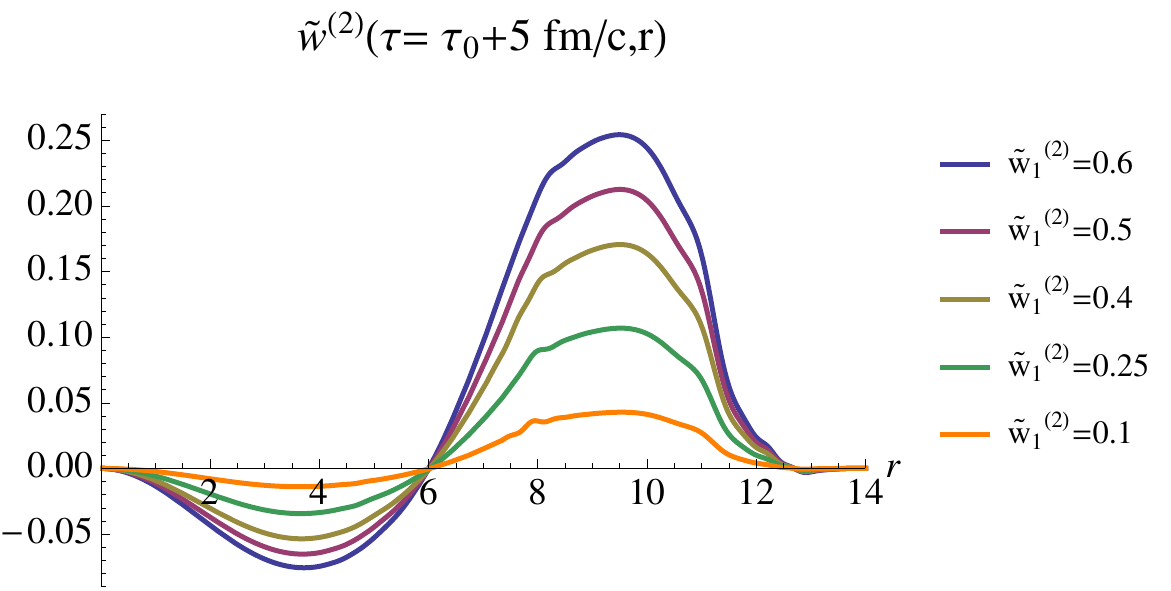}
\includegraphics[width=7.5cm]{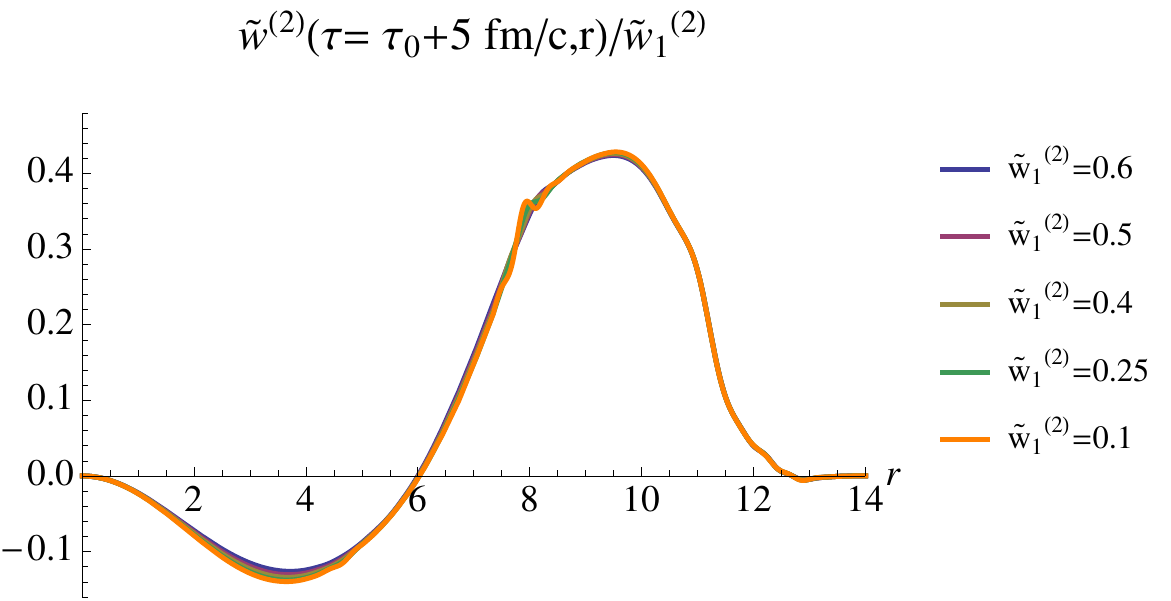}
\includegraphics[width=7.5cm]{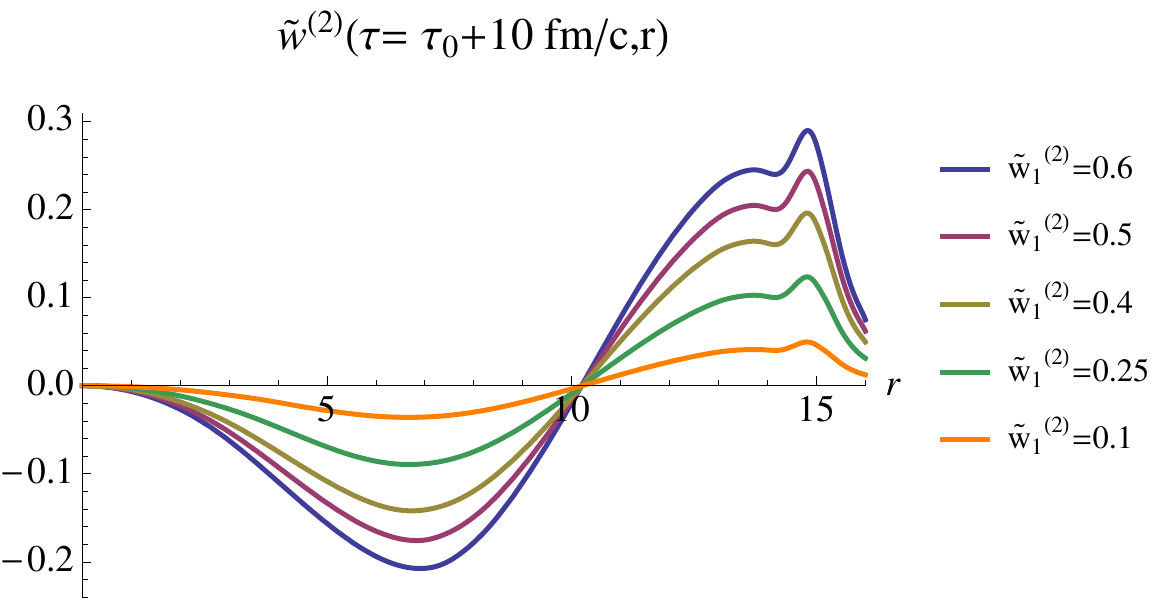}
\includegraphics[width=7.5cm]{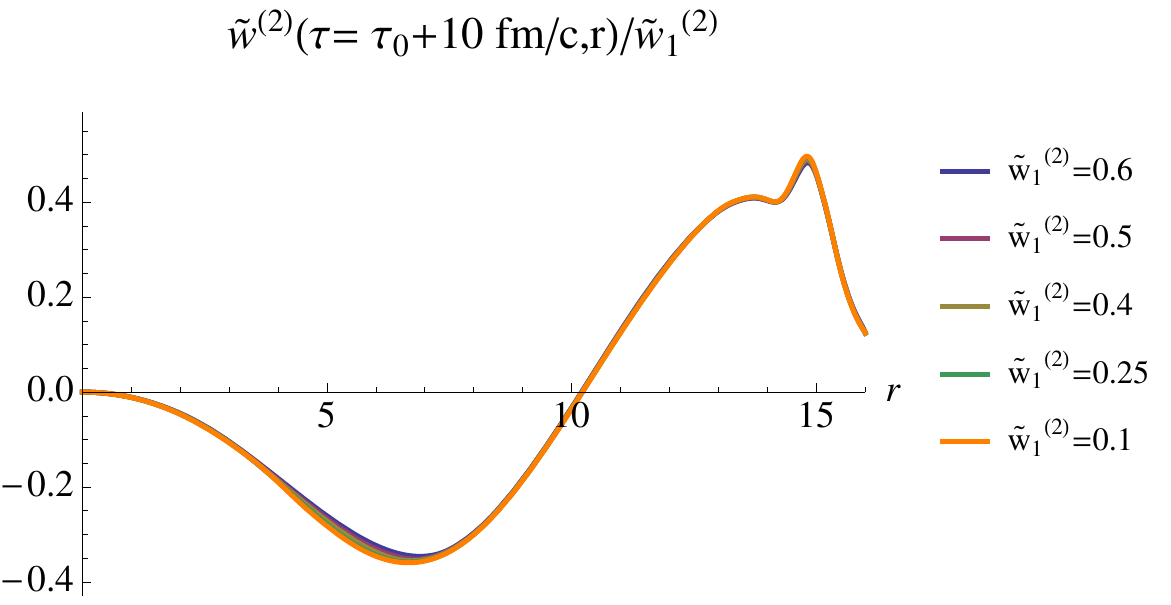}
\end{center}
\vspace{-0.5cm}
\caption{
Results for the hydrodynamic evolution of the initial condition (\ref{initial}), obtained
with ECHO-QGP. Upper row: the time dependence of the enthalpy density
is shown separately for the background $w_{BG}(\tau,r)$ and for the perturbation 
$\tilde w^{(2)}(\tau, r)$ initialized with $\tilde{w}_1^{(2)} = 0.5$.
Middle row: the dependence of the perturbations $\tilde w^{(2)}$ on the initial weight 
for $\tau=\tau_0+ 5 \text{fm/c}$ (left) and scaled by the initial weights, 
$\tilde w^{(2)}(\tau, r)/\tilde w^{(2)}_1$ (right).
This scaling establishes that the fluid dynamic response to perturbations
is approximately linear.
Lower row: same results as shown in middle row, but 
for $\tau=\tau_0+ 10 \text{fm/c}$.
}\label{fig1}
\end{figure}

For the initial conditions, we make  assumptions that are widely spread in the phenomenological literature. The initial transverse velocity components vanish, the longitudinal velocity is Bjorken boost invariant, the shear stress tensor is initialized by its Navier-Stokes value, and
the bulk viscous pressure is neglected. Initial fluctuations reside 
then only in the initial enthalpy density $w(\tau,\vec{r})$, that we parametrize in terms of an azimuthally averaged background $w_{BG}(\tau,r)$ and the weights
$\tilde w^{(m)}_l$ of the azimuthal ($m$) and radial ($l$) wave numbers of a 
discrete orthonormal Bessel-Fourier decomposition~\cite{Floerchinger:2013rya} 
\begin{equation}
	w(\tau_0, r, \varphi) = w_{BG}(\tau_0, r)\, 
	\left( 1 + \sum_{m=-\infty}^{\infty} \tilde w^{(m)}(\tau_0, r) 
	      \; e^{ i m \varphi}\right)\, ,\qquad
	      \tilde w^{(m)}(\tau_0,r) = 
	      \sum_{l=1}^\infty \tilde w^{(m)}_l\, J_{m}\left( k_l^{(m)} r \right)\, .
	      \label{eq2}
\end{equation}
Here $k^{(m)}_l = z_l^{(m)}/R$, where $z_l^{(m)}$ is the $l$-th zero of the modified Bessel function $J_m$ and $R=8$ fm throughout this work. Since $\tilde w(\tau, r,\varphi)$ is real, we
have $\tilde w^{(m)}(\tau,r) = \tilde w^{(-m)*}(\tau,r)$. In the following, we take the
weights with $m \geq 0$ as the independent ones and write
\begin{equation}
\tilde w^{(m)}_l = |\tilde w^{(m)}_l| e^{-i m \psi^{(m)}_l}\, .
\label{eq3}
\end{equation}
The corresponding modes with $m<0$ are then not independent and are defined by the condition 
$|\tilde w^{(m)}_l | = |\tilde w^{(-m)}_l |$ with azimuthal angle $\psi^{(-m)}_l = \psi^{(m)}_l \pm \pi$.

We consider first the case for which one single fluctuating basis mode is embedded 
on top of $w_{BG}(\tau_0, r)$. For example, we specify this mode with the
weight $\tilde w_1^{(2)}$, so that the initial enthalpy density reads
\begin{equation}
w(\tau_0,\vec{r}) = w_{BG}(\tau_0, r) \left[1 + 
2 |\tilde w_1^{(2)}| J_2\left( k_1^{(2)} r \right)\, 
\cos\left( 2(\varphi - \psi_1^{(2)})\right) \right]\, .
\label{initial}
\end{equation}
For one single mode, we can set without loss of generality $\psi_1^{(2)} = 0$.
Assuming for simplicity a Bjorken-boost invariant longitudinal dependence, we evolve these 
initial conditions with the 2+1 dimensional version of the hydrodynamical code ECHO-QGP~\cite{DelZanna:2013eua} with a value $\eta/s = 1/4\pi$ for the ratio of shear viscosity to entropy density~\footnote{For these $2+1$ dimensional simulations in Bjorken coordinates, we adopt a uniform grid in $x$ and $y$ with a spatial resolution of 0.2 fm,
whereas the time-step is set to $10^{-3}$ fm/c, with a Courant number
of 0.2 to ensure stability. Spatial reconstruction is achieved by employing
the MPE5 scheme, the most accurate one available in ECHO-QGP
(fifth order for smooth flows).  For further technical details, see Ref.\cite{DelZanna:2013eua}.}. Following Ref.~\cite{Qiu:2011hf}, we use the equation of state s95p-PCE
which combines lattice QCD results at high temperatures with a hadron resonance gas at low temperatures.  
The background $w_{BG}$ used throughout this paper is initialized at $\tau_0 = 0.6$ fm/c
with an azimuthally symmetric average of Glauber model initial conditions for 
Pb+Pb collisions at the LHC, described in Ref.~\cite{Floerchinger:2013vua}. 
The time evolution of $w_{BG}$ determined from ECHO-QGP is shown in Fig.~\ref{fig1}.
The time-evolved fluctuation $\tilde{w}^{(2)}(\tau, r)$ is determined from the full
hydrodynamic evolution via Fourier analysis. Results are shown in Fig.~\ref{fig1} for 
different weights $\tilde w^{(2)}_1$. Fluctuations at time
$\tau_0$ are cut-off in the region of very low background density, see e.g.
 $\tilde w^{(2)}(\tau_0,r)$ in Fig.~\ref{fig1} -- we have checked that this does 
 not affect our results. The main conclusion from Fig.~\ref{fig1} is that
 at all relevant times and even for relatively large initial amplitudes 
 $\tilde w_1^{(2)}$, the fluid dynamic response $w_{BG} \tilde w^{(2)}(\tau,\vec{r})$ 
 to an initial perturbation scales {\it approximately linearly} with the weight  
 $\tilde w_1^{(2)}$.  This is the behavior expected from the linear term in 
 eq.~\eqref{eq1}. We observe this linear dependence with similar accuracy
also for other basic modes (data not shown). 
\begin{figure}[t]
\begin{center}
\includegraphics[width=7.5cm]{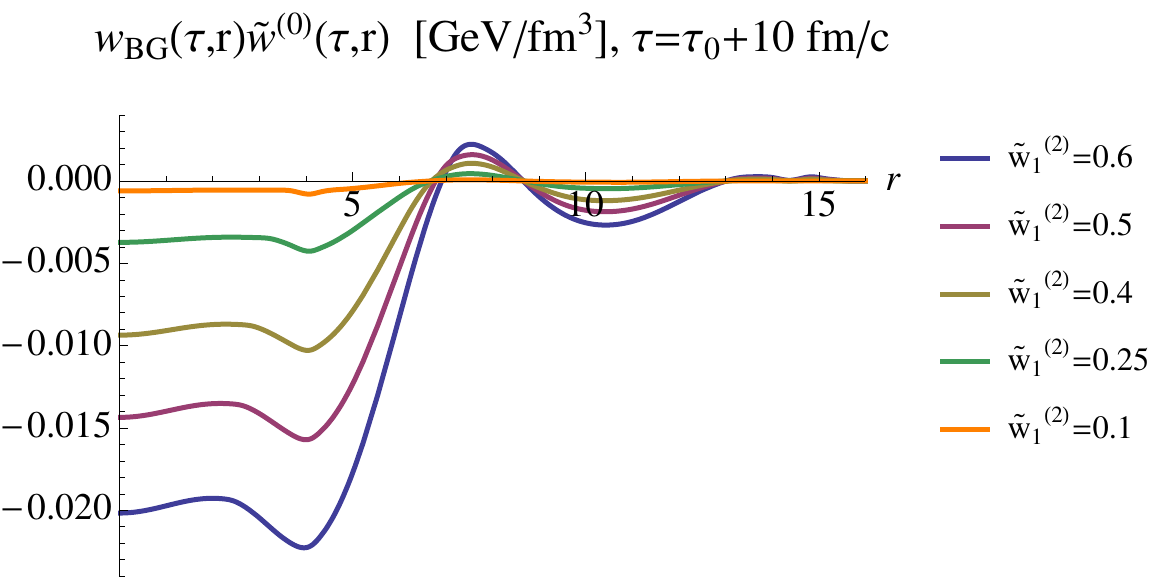}
\includegraphics[width=7.5cm]{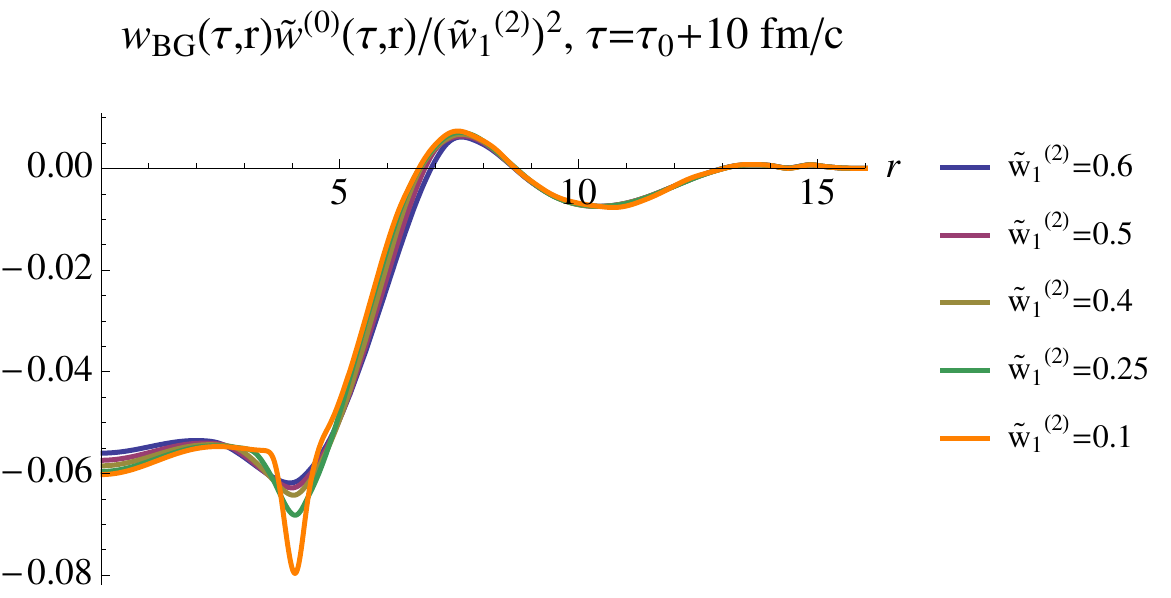}
\includegraphics[width=7.5cm]{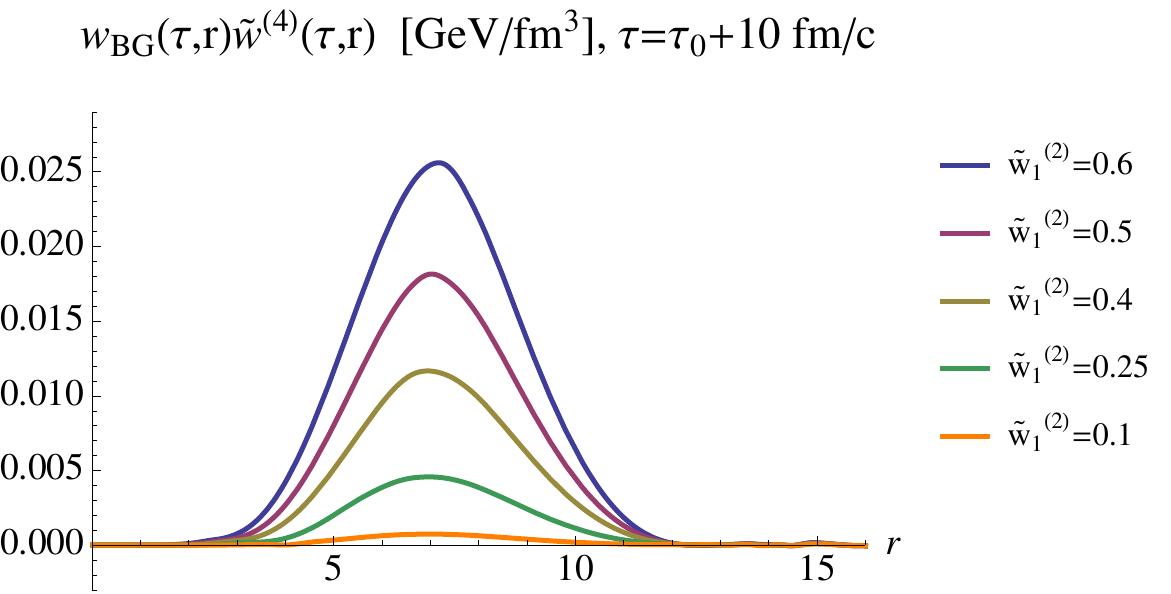}
\includegraphics[width=7.5cm]{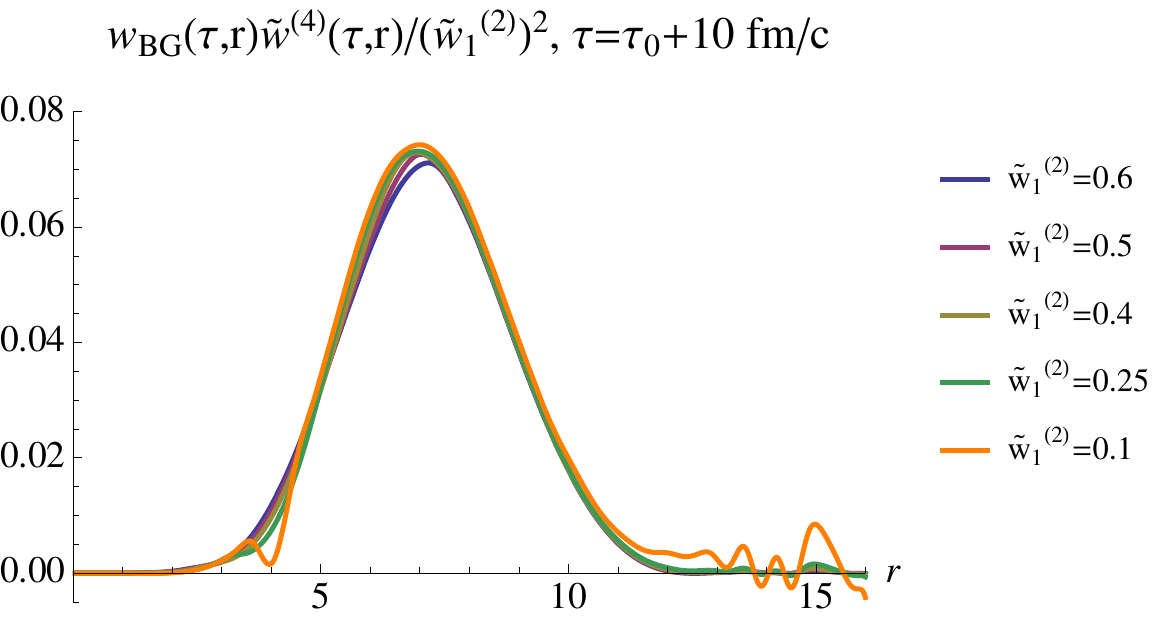}
\includegraphics[width=7.5cm]{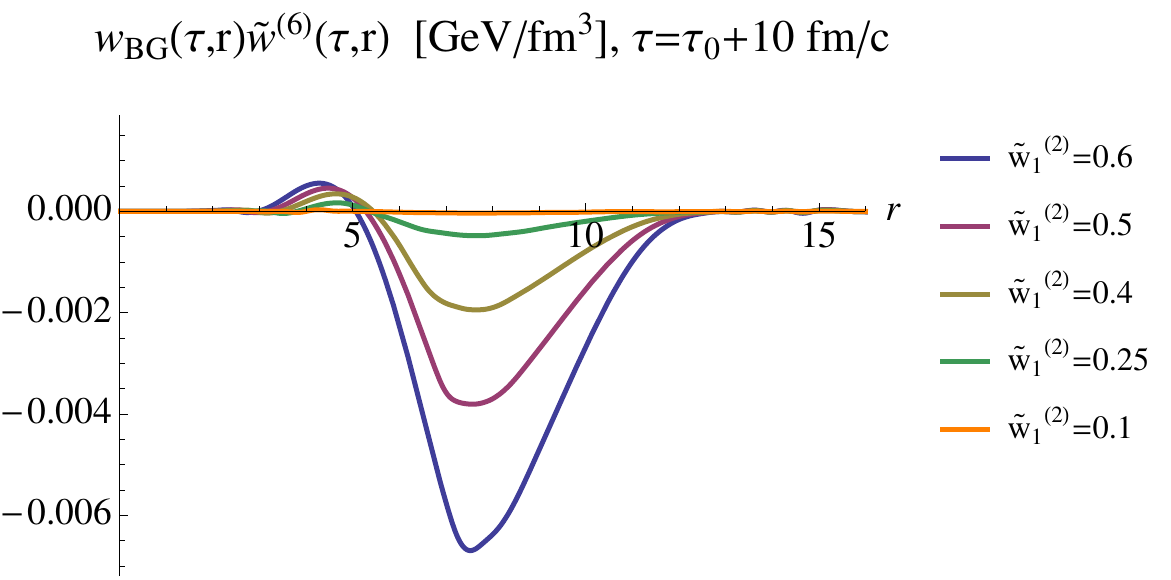}
\includegraphics[width=7.5cm]{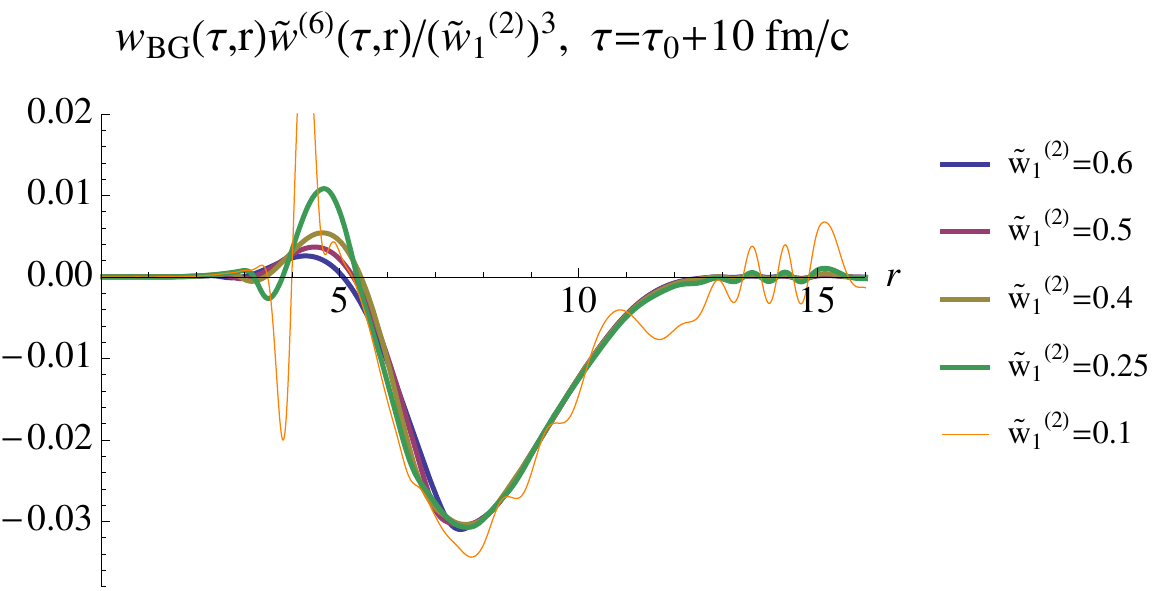}
\end{center}
\vspace{-0.5cm}
\caption{Left column: the zeroth, fourth and sixth harmonic perturbations 
induced by an initial fluctuation in the second harmonic, shown for different values of 
the initial weight $\tilde{w}_1^{(2)}$. Right column: Same results but rescaled by the
second (third) power of the weight  $\tilde{w}_1^{(2)}$. This scaling establishes that
$\tilde w^{(0)}(\tau,r)$ and $\tilde w^{(4)}(\tau,r)$ ($\tilde w^{(6)}(\tau,r)$) can be 
understood as overtones that are induced by the initial second harmonic 
perturbation as a perturbative second (third) order correction to (\ref{eq1}). 
The short-range fluctuations in the
rescaled $\tilde w^{(6)}(\tau,r)$ result from amplifying the numerical uncertainties of very
small number by a large scaling factor
$(1/\tilde w_1^{(2)})^3 = 1000$.
}\label{fig2}
\end{figure}

The almost exact linear scaling of the hydrodynamic response 
$\tilde{w}^{(m)}(r,\tau)$ with the initial weights $w^{(m)}_l$ 
does not imply that non-linearities are absent. 
To see that, consider the Fourier series 
$\tilde h_i(\tau, r, \varphi) = \frac{1}{2\pi} \sum_{m=-\infty}^\infty \; e^{im\varphi}\; \tilde h_i^{(m)}(\tau, r)$, where 
$\tilde h_i^{(m)}(\tau,r)$ are in general complex expansion coefficients, 
but $\tilde h_i^{(m)}(\tau, r) = \tilde h_i^{(-m)*}(\tau, r)$ since 
$\tilde h(\tau, r, \varphi)\in \mathbbm{R}$. Since the kernels
in (\ref{eq1}) depend only on the background field, they are invariant 
under azimuthal rotation and their Fourier expansions read
\begin{equation}
\begin{split}
\mathscr{G}_{ij}(\tau,\tau_0,r,r^\prime,\Delta\varphi) & = \frac{1}{2\pi} \sum_{m=-\infty}^\infty e^{im\Delta\varphi} \mathscr{G}_{ij}^{(m)}(\tau,\tau_0,r,r^\prime)\, ,\\
\mathscr{H}_{ijk}(\tau,\tau_0,r,r^\prime,r^{\prime\prime}, \Delta\varphi^\prime, \Delta\varphi^{\prime\prime}) & = \frac{1}{(2\pi)^2} \sum_{m^\prime,m^{\prime\prime}=-\infty}^\infty e^{i(m^\prime\Delta\varphi^\prime+m^{\prime\prime} \Delta\varphi^{\prime\prime})} \mathscr{H}^{(m^\prime,m^{\prime\prime})}_{ijk}(\tau,\tau_0,r,r^\prime,r^{\prime\prime})\, ,
\end{split}
\end{equation}
and so on. From $\mathscr{G}_{ij}(\tau,\tau_0,r,r^\prime,\Delta\varphi)\in \mathbbm{R}$ one obtains $\mathscr{G}^{(m)}_{ij} = \mathscr{G}^{(-m)*}_{ij}$ and similarly $\mathscr{H}_{ijk}^{(m^\prime,m^{\prime\prime})} = \mathscr{H}_{ijk}^{(-m^\prime,-m^{\prime\prime})*}$.
One obtains then from eq.\ \eqref{eq1}
\begin{equation}
\begin{split}
\tilde h^{(m)}_i(\tau,r) = & \int_{r^\prime} \mathscr{G}_{ij}^{(m)}(\tau,\tau_0,r,r^\prime) \, \tilde h^{(m)}_j(\tau_0,r^\prime)\\
& +\frac{1}{2} \int_{r^\prime,r^{\prime\prime}} \frac{1}{2\pi} \sum_{m^\prime,m^{\prime\prime}}\delta_{m,m^\prime+m^{\prime\prime}} \mathscr{H}_{ijk}^{(m^\prime,m^{\prime\prime})}(\tau,\tau_0,r,r^\prime, r^{\prime\prime}) \, \tilde h^{(m^\prime)}_j(\tau_0, r^\prime)\, 
\tilde h^{(m^{\prime\prime})}_k(\tau_0, r^{\prime\prime}) + \ldots
\end{split}
\label{eq:PertExpFourier}
\end{equation}
For the case that initial conditions contain only fluctuations of enthalpy density, we have
$\tilde h^{(m)}_j(\tau_0, r) = \delta_{j1} \; \tilde w^{(m)}(\tau_0, r)$. 
Using the orthonormal expansion \eqref{eq2} for $\tilde w^{(m)}(\tau_0, r)$, one can write eq.\ \eqref{eq:PertExpFourier} as
\begin{equation}
\tilde h^{(m)}_i(\tau,r) =  \sum_{l^\prime} \mathscr{G}_{i1 \, ; \, l^{\prime}}^{(m)}(\tau,\tau_0,r) \, \tilde w^{(m)}_{l^\prime} 
 +\frac{1}{4 \pi}\sum_{m^\prime,m^{\prime\prime}, l^\prime, l^{\prime\prime}}\delta_{m,m^\prime+m^{\prime\prime}} \mathscr{H}_{i11 \, ; \, l^\prime l^{\prime\prime}}^{(m^\prime,m^{\prime\prime})}(\tau,\tau_0,r) \; \tilde w^{(m^\prime)}_{l^\prime}\, 
\tilde w^{(m^{\prime\prime})}_{l^{\prime\prime}} + \ldots
\label{eq:PertExpFourierEnthalpy}
\end{equation}
with
\begin{equation}
\mathscr{G}_{i1 \, ; \, l^\prime}^{(m)}(\tau,\tau_0,r) = \int_{r^\prime} \mathscr{G}_{i1}^{(m)}(\tau,\tau_0,r,r^\prime) \; J_m\left(k^{(m)}_{l^\prime} r^\prime\right),
\end{equation} 
and similarly for $\mathscr{H}_{i11 \, ; \, l^\prime l^{\prime\prime}}$. 

According to \eqref{eq:PertExpFourier}, if one initializes fluctuations
with a {\it single} mode of weight $\tilde w_l^{(m)}$, as done in Fig.~\ref{fig1},
then corrections that are quadratic in the fluctuations $\tilde h_i(\tau_0)$  
will not appear in the time-evolved harmonics $\tilde h^{(m)}_i$, but in
the harmonics $\tilde h^{(2m)}_i$ and $\tilde h^{(0)}_i$ instead. 
Also the third order correction enters the fluctuating fields in 
$\tilde h_i^{(3m)}$ (and it enters in $\tilde h_i^{(m)}$ as a correction that is
subleading by two orders compared to the leading linear response).
To illustrate this general feature, one can compare the dominant 
linear response $w_{BG}\, \tilde{w}^{(2)}$ shown in Fig.~\ref{fig1} with the leading quadratic 
($w_{BG}\, \tilde{w}^{(0)}$, $w_{BG}\, \tilde{w}^{(4)}$) and
cubic ($w_{BG}\, \tilde{w}^{(6)}$) corrections displayed in Fig.~\ref{fig2}. 
We observe that quadratic (cubic)
corrections scale with the square (the cube) of the initial weight
$\tilde w_1^{(2)}$, as expected from \eqref{eq:PertExpFourier}. 
Moreover, even for a weight $\tilde w_1^{(2)} = 0.5$, quadratic corrections
are approximately a factor 5 smaller than the linear response, and cubic
corrections are another factor 5 smaller than the quadratic ones. 
From Ref.~\cite{Floerchinger:2013vua}, we know that for realistic initial conditions 
in heavy ion collisions,
the average weights of basis modes are of order $O(0.1)$ and
that only the tails of event distributions in  $\tilde w_l^{(m)}$ may reach
values of order 0.5. Fig.~\ref{fig2} thus indicates that non-linear corrections,
while clearly present, can be treated as small perturbations for 
fluctuations of realistic weight. 
\begin{figure}[t!]
\begin{center}
\includegraphics[width=7.5cm]{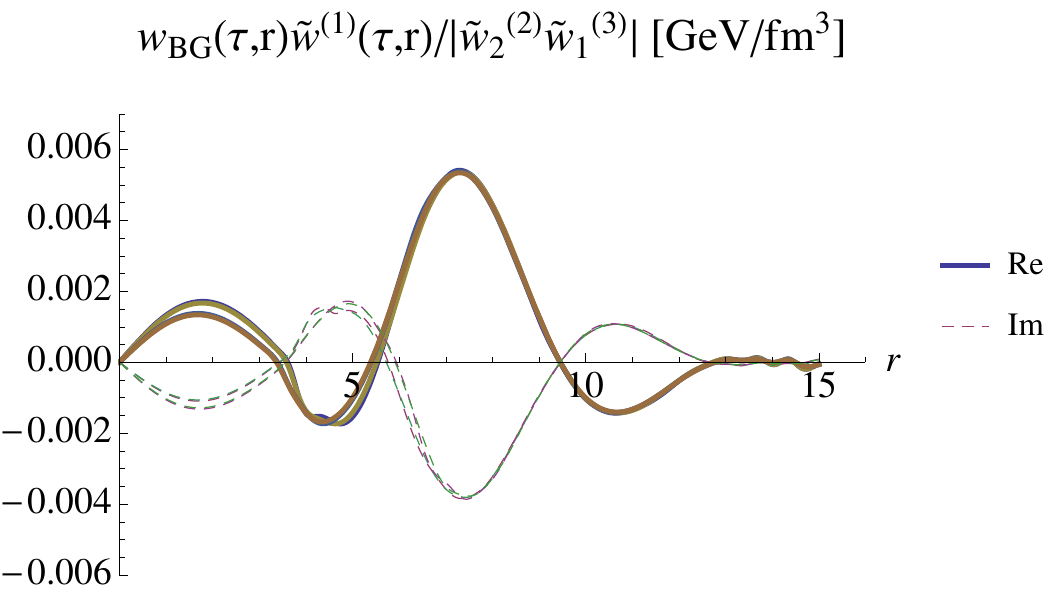}
\includegraphics[width=7.5cm]{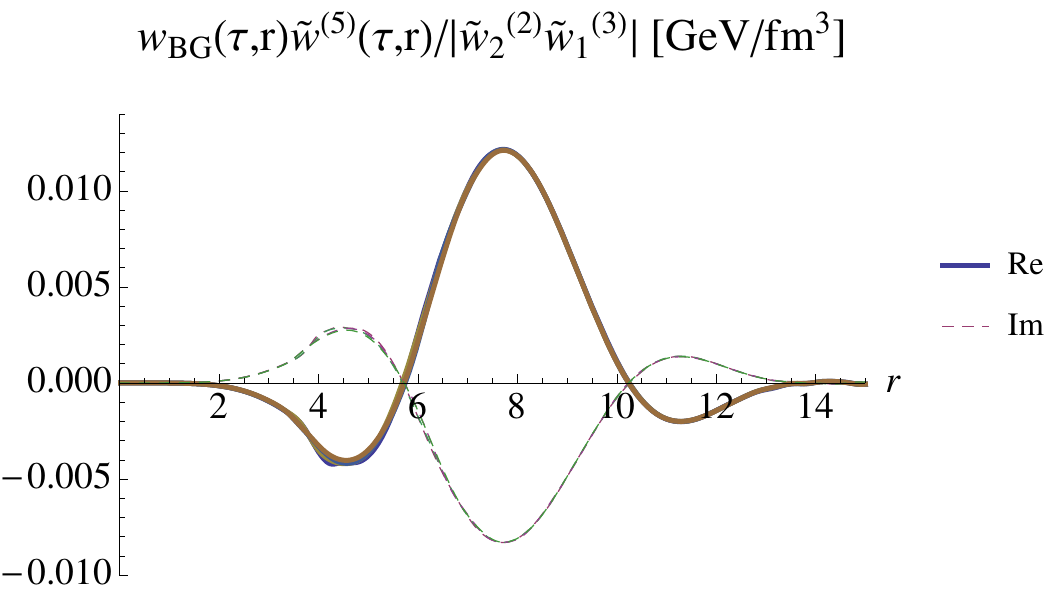}
\includegraphics[width=5.5cm]{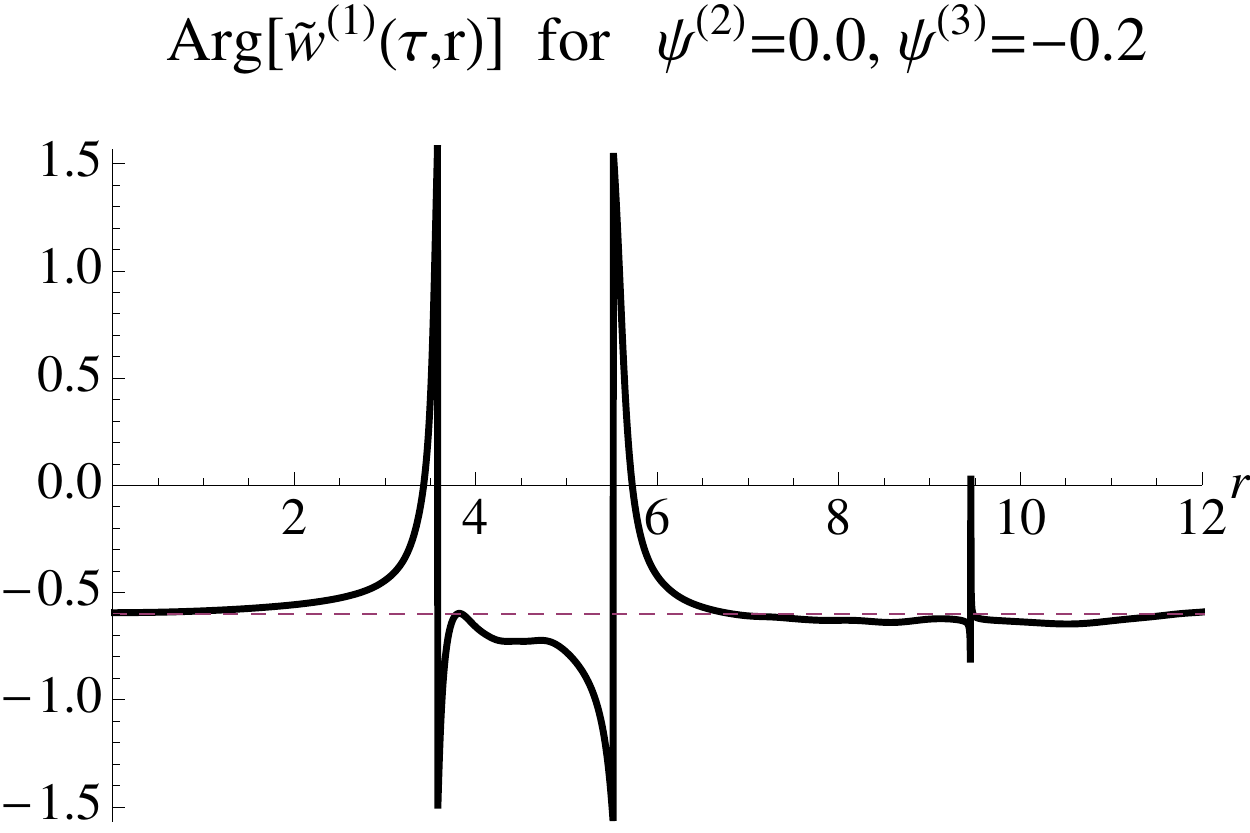}
\hspace{1.cm}
\includegraphics[width=5.5cm]{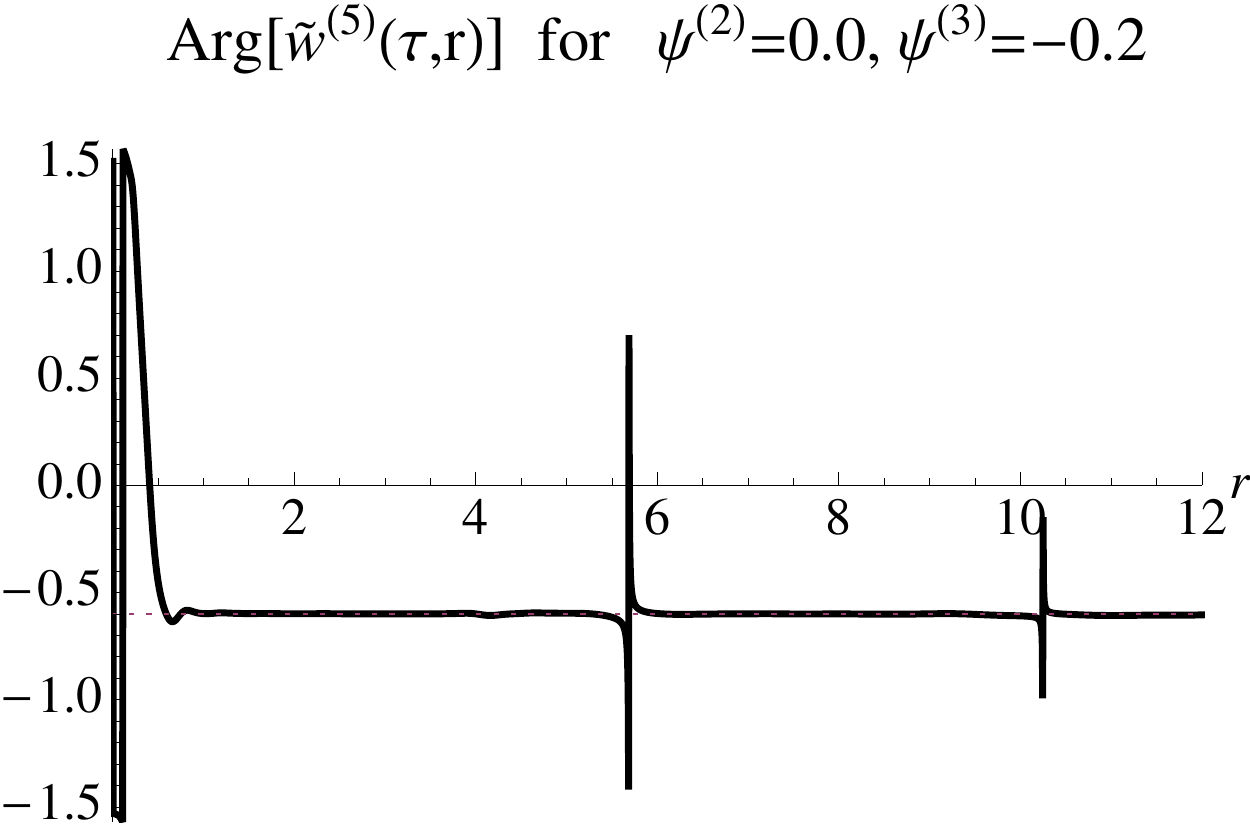}
\includegraphics[width=11.5cm]{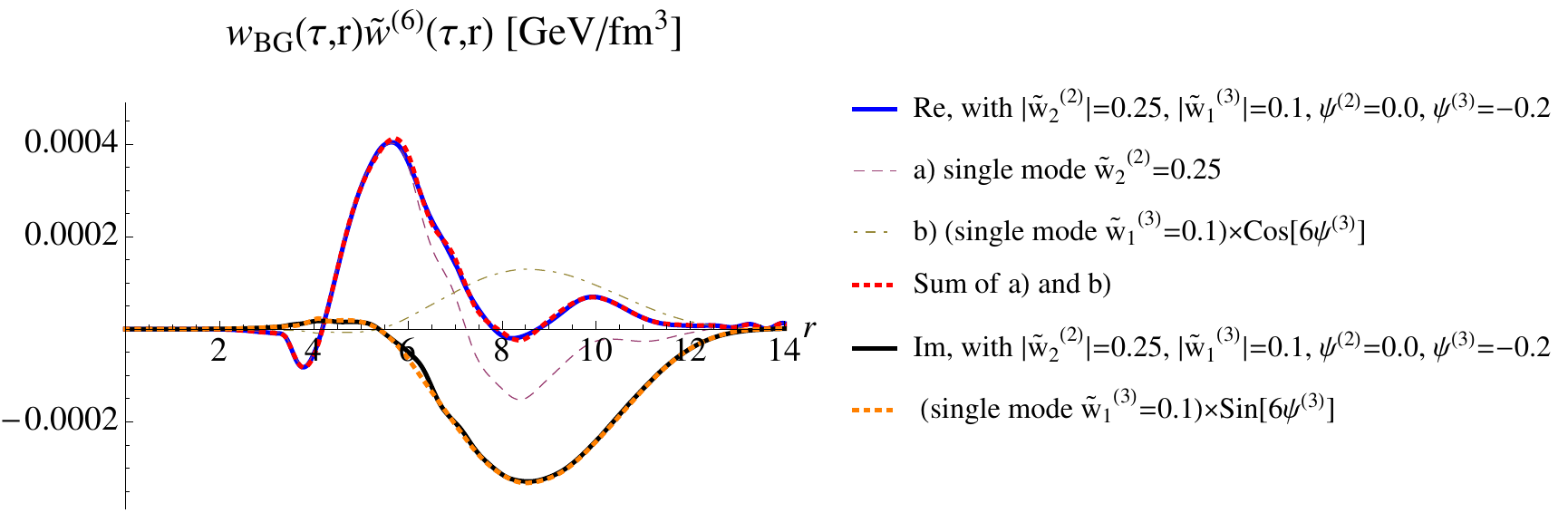}
\end{center}
\vspace{-0.5cm}
\caption{Results from ECHO-QGP for evolving up to $\tau = \tau_0 + 10$ fm/c 
on top of the background of Fig.~\ref{fig1} an initial condition composed of two basis 
modes with weights $\tilde{w}_2^{(2)} $, $\tilde{w}_1^{(3)}$ and 
angles $\psi^{(2)} = 0$ and $\psi^{(3)} = -0.2$. 
Upper row: Real and imaginary part of the first ($w_{BG}\tilde w^{(1)}$) and fifth
($w_{BG}\tilde w^{(5)}$) harmonics of the enthalpy. The curves shown are for the
four combinations of 
 $\tilde{w}_2^{(2)} = 0.1, 0.25$ and $\tilde{w}_1^{(3)} = 0.1, 0.25$ and illustrate 
 scaling behavior.
Middle row: The phase ${\rm Arg}\left[\tilde{w}^{(m)}(\tau,r)\right] $  
of the $m$-th harmonic mode (solid) compared to the perturbative expectation
(dashed line) based on Eq.~(\ref{eq:PertExpFourierEnthalpy}).
Lower row: Real and imaginary part of the sixth harmonic (solid lines). The dashed 
and dotted lines show results for the individual contributions of single basis modes.
When appropriately weighted with the phase factors according to the perturbative Eq.(\ref{eq:PertExpFourierEnthalpy}), their sum agrees with the full numerical result. 
This illustrates that the interaction between initial perturbations
of different wave numbers can be understood perturbatively.
}\label{fig3}
\end{figure}

So far, we have demonstrated with examples that 
Eq.~(\ref{eq:PertExpFourierEnthalpy}) explains the dominance of linear response
and the relative size and ordering of the overtones induced by one basis fluctuation.
We have checked extensively that the same equation explains also the
structure and symmetries of the hydrodynamic interactions between initial perturbations 
with different wave numbers. Fig.~\ref{fig3} illustrates this point with a case for which
two perturbations $\tilde{w}_2^{(2)}$,  $\tilde{w}_1^{(3)}$ are embedded on top of the
initial background fields. We have checked that the second (third) harmonics 
$\tilde{w}^{(2)}(\tau,r)$  ($\tilde{w}^{(3)}(\tau,r)$) of the fluid dynamic response
scale linearly with the initial weight $\tilde{w}_2^{(2)}$ ($\tilde{w}_1^{(3)}$) and that
they agree to high accuracy with the response to an initial configuration in which only
one mode $\tilde{w}_2^{(2)}$ ($\tilde{w}_1^{(3)}$) is embedded on top of $w_{BG}$
(data not shown). Also, $\tilde{w}^{(4)}(\tau,r)$ scales with the square of 
$\tilde{w}_2^{(2)}$ (data not shown), similarly to the case shown in Fig~\ref{fig2}.
For studying interactions between different modes, we show in Fig.~\ref{fig3} 
the first and fifth harmonics that according to eq. (\ref{eq:PertExpFourierEnthalpy}) 
are the only harmonics that receive leading second order contributions proportional to 
$\tilde{w}_2^{(2)} \tilde{w}_1^{(3)}$. If 
$\psi^{(2)} \not= \psi^{(3)}$, then the responses $w_{BG}\tilde w^{(1)}$ and
$w_{BG}\tilde w^{(5)}$ have both a real and an imaginary part.   
Both parts exhibit the expected scaling with $\tilde{w}_2^{(2)} \tilde{w}_1^{(3)}$, as
seen in Fig.~\ref{fig3}. 
Also, according to (\ref{eq:PertExpFourierEnthalpy}),
the phases of the first and fifth harmonics are determined by 
the orientations of the initial perturbations. The comparison with the full numerical results
in the middle panel of Fig.~\ref{fig3} shows that this perturbative expectation is realized approximately (strong deviations 
are seen only for values of the radius $r$ for which either 
$\text{Re} \left[ \tilde{w}^{(m)}\right]$ or $\text{Im}\left[ \tilde{w}^{(m)}\right]$ 
approach zero and for which the orientation is thus not well defined). 
To add one level of complication, we consider finally the sixth harmonics $w_{BG}\tilde w^{(6)}$ that, according to (\ref{eq:PertExpFourierEnthalpy}), 
receives corrections of second order in $\tilde{w}_1^{(3)}$ and of third order in 
$\tilde{w}_2^{(2)}$.  Fig.~\ref{fig3} shows that weighting both contributions with the
perturbatively expected information on phases and amplitude provides for a full 
quantitative understanding of the numerically determined signal $w_{BG}\tilde w^{(6)}$ 
as overtones of the two initial perturbations. 

\begin{figure}[h!]
\begin{center}
\includegraphics[width=5cm]{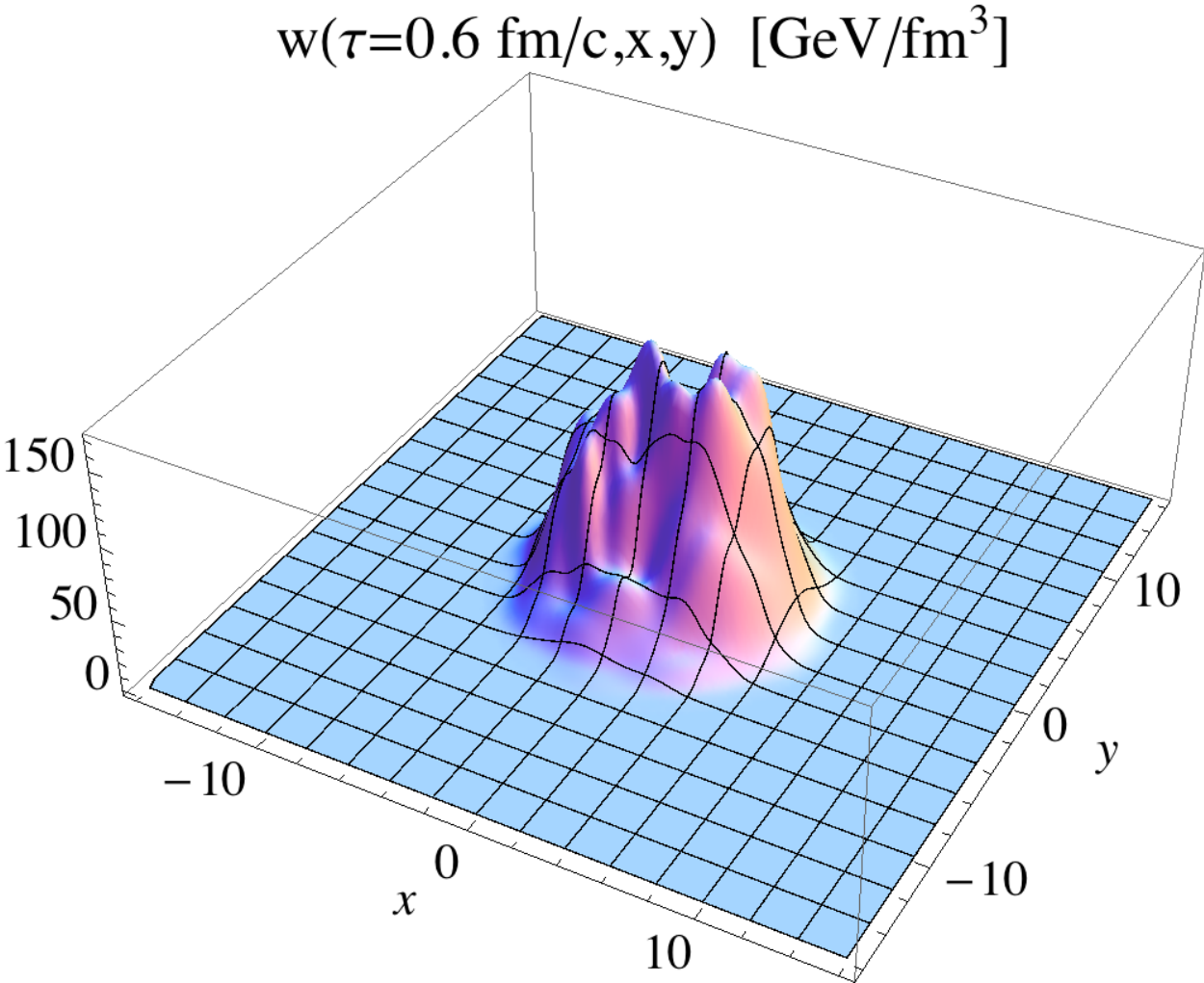}\hspace{1.cm}
\includegraphics[width=5cm]{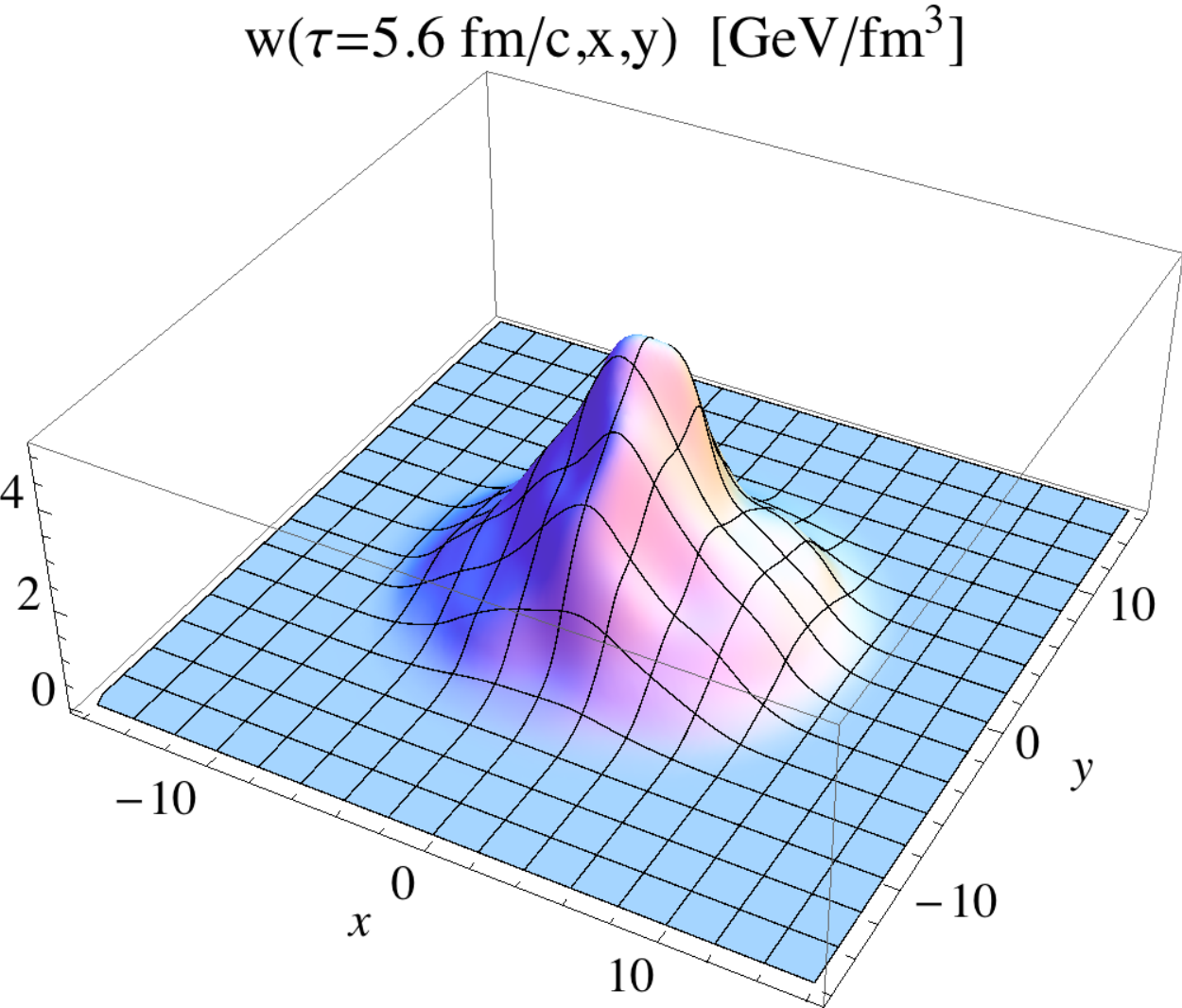}
\includegraphics[width=12.cm]{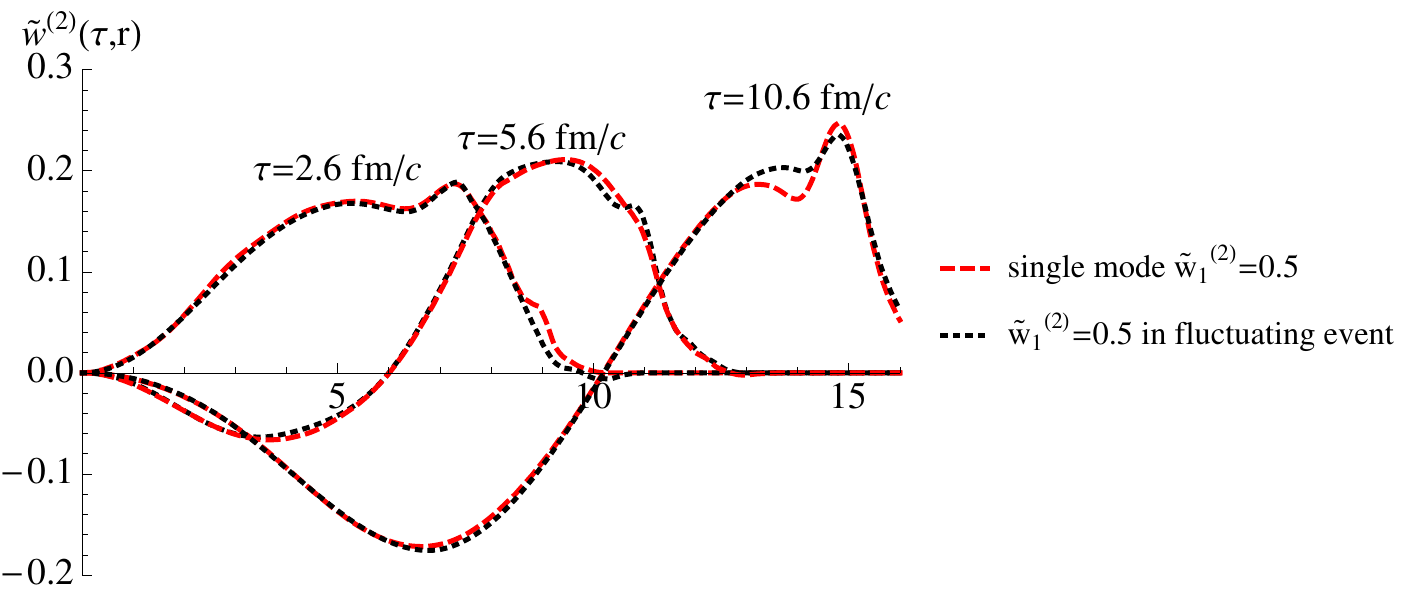}
\end{center}
\vspace{-0.5cm}
\caption{Upper panels: Left: Example of an initial condition with many 
fluctuating modes $\tilde{w}_l^{(m)}$, $m\not= 2$, and the mode $\tilde{w}_1^{(2)}=0.5$
on top of the background $w_{BG}$.
Right: The same distribution, evolved up to $\tau = \tau_0 + 5$ fm/c.
Lower panel: The second harmonics $\tilde{w}^{(2)}(\tau,r)$ extracted for 
different times $\tau$. Results extracted from the fluctuating event shown
in the upper panel are compared to the case shown in Fig.~\ref{fig1} in which 
$\tilde{w}_1^{(2)}=0.5$ is the only mode embedded on top of a smooth background.
This illustrates that the assumption of a predominantly linear response on top of a suitably
chosen background is applicable for realistic initial conditions that display strong fluctuations. }\label{fig4}
\end{figure}

Realistic initial conditions for the fluid dynamic evolution of heavy ion collisions are expected 
to involve fluctuations on many different length scales and with large amplitude.
Could it be that the examples discussed so far, although initialized with relatively
large amplitudes,  are still academic, and that they cannot be extended to deal with the 
complexity of a realistic heavy ion collision? To lay such concerns at rest, we have 
embedded simple basis modes in realistic initial conditions with many and large fluctuations. 
Fig.~\ref{fig4} shows such an initial distribution. It is constructed by subtracting from
an arbitrary initial condition generated by a Glauber model the contribution leading to a 
second harmonic and adding then the perturbation of (\ref{initial}). In this way, we have
an analytically controlled initial perturbation on top of a realistically fluctuating background,
and we can extract this initial perturbation and the time dependence of 
its fluid dynamic response via Fourier analysis. The lower panel of Fig.~\ref{fig4} shows
that this dynamical response in an event with realistic fluctuations 
is described to high accuracy by the linear response 
on top of the smooth background that we had established in Fig.~\ref{fig1}. 
 
In summary, the evolution of initial anisotropic density perturbations as determined numerically with the fluid dynamics solver ECHO-QGP seems to follow a simple pattern that can be understood order-by-order in a perturbative expansion for small deviations from an azimuthally symmetric event-averaged background. The leading order is linear and modes with different azimuthal wave numbers do not mix. Quadratic and higher orders can be seen as next-to-leading order corrections. They influence modes with azimuthal wave numbers that can be written as sums (or differences) of the seed wave numbers. Since the non-linear couplings are numerically small, the higher harmonics generated by two-mode or three-mode interactions will often be small in comparison to initially present and linearly evolving perturbations. This ordering may be less pronounced for non-central collisions where the elliptic modes (with $m=2$) have a particularly strong weight such that its overtones with $m=4, 6$ etc.\ may dominate over primordial density perturbations with these wave numbers. We also note that the relative importance of linear and non-linear terms depends significantly on the dissipative properties of the medium. In exploratory studies that lie outside the scope of this letter, we found that for increasing
$\eta/s$, the linear response is more dominant and the relative weight of higher 
non-linear orders decreases. Such a more laminar behavior is expected on general
grounds. All results shown in this Letter are for a rather minimal value of $\eta/s = 1/4\pi$
which maximizes contributions from non-linear corrections. 

The perturbative picture expressed by eqs.\ \eqref{eq1} or \eqref{eq:PertExpFourier} provides a useful ordering scheme for a more detailed understanding of the evolution of fluctuations in hydrodynamic fields. The numerical results discussed here provide motivation for a more formal and thorough development of this kind of perturbation theory. It is unclear whether all initial perturbations in hydrodynamic fields follow a similar pattern. In particular, from general fluid dynamic considerations one may expect that non-linear terms are more important for vorticity excitations \cite{Florchinger:2011qf}.

Another open question concerns the freeze-out from hydrodynamic fields to particle spectra. To linear order in a mode-by-mode treatment of fluid dynamic perturbations this was discussed recently \cite{Floerchinger:2013hza}. However, non-linear corrections arise there as well. For instance, the proportionality 
$v_4 \propto (v_2)^2$ was conjectured in \cite{Borghini:2005kd} as a consequence of this effect. It would be interesting to see how large these non-linearities are in comparison to the ones from the fluid dynamic propagation as discussed here.

It will also be interesting to investigate whether the approach
presented here can provide a more detailed dynamical underpinning 
of results in the recent literature. For instance, the experimentally
accessible reaction plane correlations formulated and studied in
Refs.~\cite{Bhalerao:2011yg,Qin:2011uw,Jia:2012ju,Teaney:2013dta} are akin of
the constraints on the phases of higher order contributions
in the perturbative  ansatz (\ref{eq:PertExpFourierEnthalpy}). Similarly, the (non-linear) response to cumulants of the initial density distribution as formulated and studied in Refs.~\cite{Teaney:2010vd,Teaney:2012ke,Teaney:2013dta} may be related to some integrated version of eq.\ (\ref{eq:PertExpFourier}).

\paragraph{Acknowledgements}
AB is supported by European Research Council grant HotLHC ERC-2011-StG-279579.
ECHO-QGP was developed within the Italian PRIN 2009 MIUR project
"Il Quark-Gluon Plasma e le collisioni nucleari di alta energia". We also thank
F. Becattini for helping to make this collaboration possible.


\vskip -.5cm
\begin{thebibliography}{}
%
\bibitem{Heinz:2013th}
  U.~W.~Heinz and R.~Snellings,
  arXiv:1301.2826 [nucl-th].
%

\bibitem{Gale:2013da}
  C.~Gale, S.~Jeon and B.~Schenke,
  Int.\ J.\ Mod.\ Phys.\ A {\bf 28} (2013) 1340011.

\bibitem{Hippolyte:2012yu}
  B.~Hippolyte and D.~H.~Rischke,
  Nucl.\ Phys.\ A {\bf 904-905} (2013) 318c
  [arXiv:1211.6714 [nucl-ex]].

\bibitem{Teaney:2009qa}
  D.~A.~Teaney,
  review for 'Quark Gluon Plasma 4', eds. R.C. Hwa and
  X.N. Wang, World Scientific, Singapore, arXiv:0905.2433 [nucl-th].

\bibitem{Qiu:2011iv}
  Z.~Qiu and U.~W.~Heinz,
  Phys.\ Rev.\ C {\bf 84} (2011) 024911.
  
\bibitem{Schenke:2011bn}
  B.~Schenke, S.~Jeon and C.~Gale,
  Phys.\ Rev.\ C {\bf 85} (2012) 024901.

\bibitem{Bhalerao:2011yg}
  R.~S.~Bhalerao, M.~Luzum and J.~-Y.~Ollitrault,
  Phys.\ Rev.\ C {\bf 84} (2011) 034910.
  
  
\bibitem{Schenke:2012wb}
  B.~Schenke, P.~Tribedy and R.~Venugopalan,
  Phys.\ Rev.\ Lett.\  {\bf 108} (2012) 252301.

\bibitem{Holopainen:2010gz}
  H.~Holopainen, H.~Niemi and K.~J.~Eskola,
  Phys.\ Rev.\ C {\bf 83} (2011) 034901.
 
\bibitem{Teaney:2010vd}
  D.~Teaney and L.~Yan,
  Phys.\ Rev.\ C {\bf 83} (2011) 064904.

\bibitem{Teaney:2012ke}
  D.~Teaney and L.~Yan,
  Phys.\ Rev.\ C {\bf 86} (2012) 044908.

\bibitem{Gardim:2011xv}
  F.~G.~Gardim, F.~Grassi, M.~Luzum and J.~-Y.~Ollitrault,
  Phys.\ Rev.\ C {\bf 85} (2012) 024908.

\bibitem{Petersen:2012qc}
  H.~Petersen, R.~La Placa and S.~A.~Bass,
  J.\ Phys.\ G {\bf 39} (2012) 055102.

\bibitem{Qian:2013nba}
  W.~-L.~Qian, P.~Mota, R.~Andrade, F.~Gardim, F.~Grassi, Y.~Hama and T.~Kodama,
  arXiv:1305.4673 [hep-ph].

\bibitem{Niemi:2012aj} 
  H.~Niemi, G.~S.~Denicol, H.~Holopainen and P.~Huovinen,
  Phys.\ Rev.\ C {\bf 87}, 054901 (2013)
  [arXiv:1212.1008 [nucl-th]].
  
  
\bibitem{Deng:2011at}
  W.~-T.~Deng, Z.~Xu and C.~Greiner,
  Phys.\ Lett.\ B {\bf 711} (2012) 301.
  
\bibitem{Floerchinger:2013rya}
  S.~Floerchinger and U.~A.~Wiedemann,
  Phys.\ Lett. B {\bf 728}, 407 (2014).
  [arXiv:1307.3453]
%

\bibitem{Floerchinger:2013vua}
  S.~Floerchinger and U.~A.~Wiedemann,
  Phys.\ Rev.\ C {\bf 88} (2013) 044906
  [arXiv:1307.7611 [hep-ph]].




\bibitem{Qiu:2011hf}
  Z.~Qiu, C.~Shen and U.~Heinz,
  Phys.\ Lett.\ B {\bf 707}, 151  (2012).

    


     
 

 

  
\bibitem{Schukraft:2012ah}
  J.~Schukraft, A.~Timmins and S.~A.~Voloshin,
  Phys.\ Lett.\ B {\bf 719} (2013) 394.
  

\bibitem{Floerchinger:2013hza}
  S.~Floerchinger and U.~A.~Wiedemann,
  arXiv:1311.7613 [hep-ph].

\bibitem{DelZanna:2013eua}
  L.~Del Zanna, V.~Chandra, G.~Inghirami, V.~Rolando, A.~Beraudo, A.~De Pace, G.~Pagliara and A.~Drago {\it et al.},
  Eur.\ Phys.\ J.\ C {\bf 73} (2013) 2524
  [arXiv:1305.7052 [nucl-th]].
    
\bibitem{Borghini:2005kd}
  N.~Borghini and J.~-Y.~Ollitrault,
  Phys.\ Lett.\ B {\bf 642} (2006) 227
  [nucl-th/0506045].
    
\bibitem{Florchinger:2011qf}
  S.~Floerchinger and U.~A.~Wiedemann,
  JHEP {\bf 1111} (2011) 100
  [arXiv:1108.5535 [nucl-th]].
  
\bibitem{Qin:2011uw}
  G.~-Y.~Qin and B.~Muller,
  Phys.\ Rev.\ C {\bf 85} (2012) 061901
  [arXiv:1109.5961 [hep-ph]].

\bibitem{Jia:2012ju}
  J.~Jia and D.~Teaney,
  Eur.\ Phys.\ J.\ C {\bf 73} (2013) 2558
  [arXiv:1205.3585 [nucl-ex]].
  
\bibitem{Teaney:2013dta}
  D.~Teaney and L.~Yan,
  arXiv:1312.3689 [nucl-th].
  
\end{thebibliography}
\end{document}